\documentclass[sigconf]{acmart}
\renewcommand\footnotetextcopyrightpermission[1]{} 
\usepackage{latexsym}
\usepackage{color}
\usepackage[linesnumbered,ruled,vlined]{algorithm2e}
\usepackage{array}

\usepackage{enumitem}
\setitemize{leftmargin=*}

\newcommand{\PreserveBackslash}[1]{\let\temp=\\#1\let\\=\temp}
\newcolumntype{C}[1]{>{\PreserveBackslash\centering}p{#1}}
\newcolumntype{R}[1]{>{\PreserveBackslash\raggedleft}p{#1}}
\newcolumntype{L}[1]{>{\PreserveBackslash\raggedright}p{#1}}

\usepackage{array,multirow,graphicx}
\newcommand{\STAB}[1]{\begin{tabular}{@{}c@{}}#1\end{tabular}}

\def\BibTeX{{\rm B\kern-.05em{\sc i\kern-.025em b}\kern-.08emT\kern-.1667em\lower.7ex\hbox{E}\kern-.125emX}}
    
\copyrightyear{2019}
\acmYear{2019}
\setcopyright{none}

\newcommand{\writes}[1]{\ensuremath{{\it write\_locs}(#1)}}
\newcommand{\instr}[1]{\ensuremath{{\it instr\_locs}(#1)}}

\newtheorem{auxdefn}{Definition}[section]

\begin{document}

%
\title{Representing and Reasoning about Dynamic Code}

\author{Jesse Bartels}
\email{jessebartels@email.arizona.edu}
\author{Jon Stephens}
\email{stephensj2@email.arizona.edu}
\author{Saumya Debray}
\email{debray@email.arizona.edu}
\affiliation{%
  \institution{University Of Arizona}
  \city{Tucson}
  \state{Arizona}
}

\renewcommand{\shortauthors}{Jesse Bartels, John Stephens, Saumya Debray}

%
\begin{abstract}
Dynamic code, i.e., code that is created or modified at runtime, is ubiquitous
in today's world.  The behavior of 
dynamic code can depend on the logic of the dynamic code generator
in subtle and non-obvious ways, with significant security implications,
e.g., JIT compiler bugs can lead to exploitable vulnerabilities in the
resulting JIT-compiled code.  Existing approaches to program 
analysis do not provide adequate support for reasoning about
such behavioral relationships.  This paper takes a first step in 
addressing this problem by describing
a program representation and a new notion of dependency that allows us to
reason about dependency and information flow relationships between the dynamic
code generator and the generated dynamic code.
Experimental results show that analyses based on these concepts are able to 
capture properties of dynamic code that cannot be identified using traditional
program analyses.
\end{abstract}

%

\maketitle

\section{Introduction}\label{sec:intro}
Dynamic code, i.e., code that is created or modified at runtime, is ubiquitous
in today's world.  Such code arises in many contexts, including JIT-compilation,
dynamic binary translation, obfuscation, and dynamic code unpacking in malware.
Dynamic code raises new security challenges, arising partly from the fact that 
the behavior of an application containing dynamic code may depend in part on
logic that is not part of the application itself, but rather is in the dynamic 
code generator.  As a concrete example, Rabet describes a JIT compiler bug in
Chrome's V8 JavaScript engine that causes some initialization
code in the application program to be (incorrectly) optimized away,
resulting in an exploitable vulnerability (CVE-2017-5121) \cite{rabet2017}.
As another example, Frassetto {\em et al.} describe how a memory corruption 
vulnerability can be used to modify the byte code of an interpreted program such 
that subsequent JIT compilation results in the creation of the malicious 
payload \cite{frassetto2017jitguard}.  As these examples suggest, 
vulnerabilities arising from dynamic code generation pose a significant security
challenge \cite{song2015exploiting,park2018bytecode}.  
To deal with such situations,
it would be helpful to be able to start from some appropriate point in the
dynamically generated code---e.g., an instruction that crashes as a result of the bug, or 
is part of the exploit code---and trace dependencies back, into and through the
JIT compiler's code, to help identify the bug that caused incorrect code to
be generated or the byte code to be corrupted.  The key here is to be able to
reason about dependencies and information flow between the dynamic
code generator and the dynamically generated code so that we can try to
determine how the behavior of the generated code is influenced by that of
the code generator. Such analyses, which we refer to as {\em end-to-end analyses},
can significantly speed up the process of identifying and fixing such problems.
Other examples of security issues arising from dynamic code, which can also
benefit from end-to-end analysis, include evasion based on environmental
triggers, where the dynamic code generator
generates different code based on different environmental values. 

Unfortunately, existing approaches to (static or dynamic) program analysis do not
adequately support such reasoning about dynamic code modification.
Traditional program representations, such as control flow graphs, 
cannot handle the effects of runtime changes to the code, which
require accommodating the possibility of some memory locations having
different instructions at different times during execution.
JIT compilers \cite{Tracemonkey-JIT,Inoue2012} and dynamic binary translators
\cite{valgrind} maintain representations of the code being
dynamically modified, but not together with that of the code that performs 
code modification.
Whole-system analyses \cite{yin2007panorama,yin2010temu,henderson2014make,%
chow2004understanding,enck2014taintdroid} perform
dynamic taint propagation, taking into account explicit information
flows via data dependencies but not implicit flows via control dependencies.
As we discuss later, they also do not take into account dependencies
that can arise through the act of dynamic code modification.
Thus, existing approaches to reasoning about program behaviors
suffer from the following shortcomings:
\begin{enumerate}
\item[$(a)$]
they do not provide program representations that allow us to even ask questions
such as ``{\em What logic in the dynamic code generator affected the generation
of the faulty application code?}''; and 
\item[$(b)$]
they do not support notions of dependence that can allow us to reason about the
computation in ways that can help answer such questions.
\end{enumerate}
This paper shows how this problem can be addressed via
a program representation that is able to
capture the structure and evolution of code that can change dynamically,
together with a notion of dependency that arises from the process of 
dynamic code generation and which is not captured by conventional notions of data
and control dependencies.  Experimental results show that our ideas make it
possible to reason about dynamic code in novel ways, e.g., we can
construct backward dynamic program slices, starting from incorrect dynamically 
generated JIT-compiled code, to include the JIT-compiler logic responsible for
the problem; and detect situations where a dynamic code generator embeds 
environmental triggers in dynamically generated code.
Such end-to-end analyses are not possible using current approaches to program analysis.

\section{Background: Control flow graphs}\label{sec:motivation}\label{sec:CFGs}

Program analyses are based on representations of the program's structure;
for concreteness, we focus on control flow graphs (CFGs).  
CFG construction for static code via static analysis is well-understood
\cite{RedDragon}.  However, this approach is inadequate for dynamic code because 
code created at runtime is not available for static inspection; instead, we use 
dynamic analysis.  This has the benefit of being able to handle dynamic code; its 
drawback is that the constructed CFG may not contain all of the program's code
due to incomplete code coverage.
We sketch here how CFGs for static code can be constructed from an instruction 
trace obtained via dynamic analysis. The extension of this approach to 
dynamic code is discussed in Section \ref{sec:pgm-repr}.

Let $G$ denote the CFG under construction. We process instructions in the execution
trace as they are encountered.  
For each instruction $I$, its properties (e.g.,
whether or not it is a control transfer) and its status within $G$ (e.g., whether
or not it is already in $G$) determine how it is processed; we refer to
this as ``{\em processing I in the context of G}.''
If $I$ has not been encountered previously, it is
added as a new instruction.  If $I$ follows a conditional or unconditional
jump, it should begin a basic block: thus, if $I$ is currently in $G$ and is
not the first instruction of its block, the block has to be split and control
flow edges added appropriately.  

Multi-threading introduces additional complexity because adjacent instructions in
the execution trace may be from different threads and thus may not represent
adjacent instructions in the code.  We handle this by tracking the construction of 
$G$ across all threads, with each update happening according to the 
state of the current thread. By state, we mean that we separately maintain the call 
stack, previous instruction seen, current function being reconstructed, etc. for each 
thread (i.e. the last instruction from one thread may be appending an instruction to 
a basic block whereas a different thread could be splitting a different block).

\section{Reasoning about Dynamic Code}\label{sec:approach}

Dynamic code modification can give rise to different versions of 
the program, with different instructions and behaviors, at different
points in its execution.
A representation suitable for
end-to-end analysis of dynamic code should keep track of the different
versions of the code resulting from dynamic modification.  There are two
issues to consider here: (1) what constitutes ``dynamic code modification''?
and (2) how should such modifications be captured in the program representation?  We
address these questions as follows.
First, we note that in general, heuristic approaches, such as categorizing
a memory write as code modification if it targets an executable
section of the program's memory, may not be sufficiently precise, e.g., because 
permissions on memory pages can be changed during execution, making a non-executable
memory region executable.  We therefore consider a write to a memory location $\ell$ as
``code modification'' only if  $\ell$ is part of some instruction that is subsequently
executed.  Second, even small dynamic code modifications can result in arbitrarily large
changes to the program's representation and behavior.  In the x86 ISA, for example, the
arithmetic instruction ``bitwise or'' (opcode: {\em or}; encoding: \texttt{0x0c}) can,
by flipping a single bit, be changed to the control transfer instruction ``jump if equal''
(opcode: {\em je}; encoding: \texttt{0x0f}).

Motivated by these observations, we build our program's CFG using
dynamic analysis, as described in Section \ref{sec:CFGs}, until we encounter
an instruction whose memory locations have been modified.  At this point we are
confronted with a potentially arbitrary change to the program's behavior and
representation.  To capture this, we begin construction of a new CFG, which we
link to the previously constructed CFG using a special type of edge that we
call a ``dynamic edge.''  Each such linked CFG corresponds to a ``phase'' of the
program's execution.  We make this notion more precise below.

\paragraph{Terminology.}
In some situations, it may make sense to distinguish between
code created at runtime prior to being executed (``dynamic code generation'')
and code modified at runtime after it has already been executed
(``dynamic code modification'').  The ideas described here apply to both
these situations, and we use the terms ``generation'' and ``modification'' 
of dynamic code interchangeably.

\subsection{Concepts and Definitions}

\begin{figure}
\begin{center}
\includegraphics[width=0.6\linewidth]{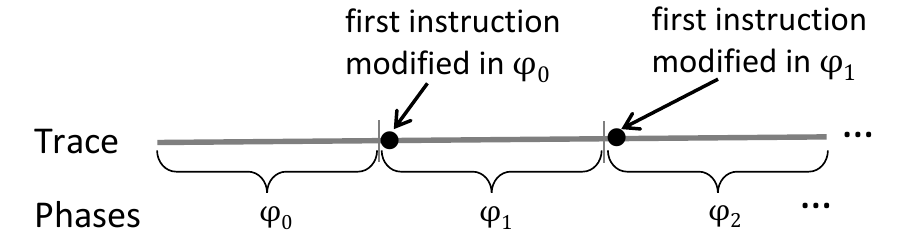}
\end{center}
\caption{Phases}\label{fig:phases}
\end{figure}

\subsubsection{Phases}
The idea behind phases is to partition an execution
of a program into a sequence of fragments
$\varphi_0, \varphi_1, \ldots, \varphi_i, \ldots$ 
such that for each $\varphi_i$, none of the locations 
written by the instructions in $\varphi_i$ is part of any instruction executed
by $\varphi_i$.  Each $\varphi_i$ is referred to as a {\em phase}.
Execution begins in phase $\varphi_0$ with the program's initial code.
When the first dynamic instruction 
is encountered, we switch to $\varphi_1$.  Execution 
continues in $\varphi_1$ (including other instructions that may have been 
created or modified in $\varphi_0$) until an instruction is encountered that
was modified in $\varphi_1$, at which point we switch to $\varphi_2$, and so on.
This is illustrated in Figure \ref{fig:phases}.

More formally, given a dynamic instance $I$ of
an instruction in a program, let \instr{I} denote the set of locations
occupied by $I$ and \writes{I} the set of locations written by $I$.  These
notions extend in a straightforward way to a sequence of instructions $S$:
\begin{quote}
$\instr{S} = \bigcup_{I \in S} \instr{I}$\\
$\writes{S} = \bigcup_{I \in S} \writes{I}$
\end{quote}
Given an execution trace $T$ for a program, let $T[i]$ denote
the $i^{th}$ instruction in $T$, and $T[i:j]$ denote the sequence (subtrace)
$T[i], \ldots, T[j]$.  We define the phases of $T$ as follows:
\begin{definition}
Given an execution trace $T$, the phases of $T$, denoted $\Phi(T)$, is a
sequence $\varphi_0, \varphi_1, \ldots, \varphi_i, \ldots$ of subtraces of $T$ 
such that the following hold:
\begin{itemize}
\item
\begin{tabbing}
$\varphi_0 = T[0$\=$:k]$, where $k = \max \{j \mid $XX\= \kill
$\varphi_0 = T[0:k]$, where
$k = \max \{\/j \mid j \geq 0$ and\\
\>     $\writes{T[0:j]} \bigcap \instr{T[0:j]} = \emptyset\};$
\end{tabbing}

\item
For $i \geq 0$, let $\varphi_i = T[k:(m-1)]$, then
\begin{tabbing}
$\varphi_0 = T[0$\=$:k]$, where $k = \max \{j \mid $XX\= \kill
$\varphi_{i+1} = T[m:n]$, where
$n = \max \{\/j \mid j \geq m$ and\\\
\>    $\writes{T[m:j]} \bigcap \instr{T[m:j]} = \emptyset\}.$
\end{tabbing}
\end{itemize}
\end{definition}
A program execution's phases capture the different ``versions'' of 
code arising from dynamic code modification during that execution.  
An execution with no dynamic code consists of a single phase.  

\subsubsection{Dynamic Control Flow Graphs}\label{sec:DCFG}
We use the notion of phases to construct control flow graphs for dynamic
code: we construct a CFG for each phase of the execution,
as discussed in Section \ref{sec:CFGs}, and link them together using
special edges, called {\em dynamic edges}, that represent the control 
flow from the last instruction of one phase to the first instruction of
the next phase.  We refer to such a CFG as a {\em dynamic control flow graph}
(DCFG).  More formally:

\begin{definition}
Given an execution trace $T$ for a program, let $\Phi(T) =
\varphi_0, \ldots, \varphi_n$ denote the phases of $T$, and
let $G_i = (V_i, E_i)$ denote the CFG constructed from the subtrace $\phi_i$.
Then the dynamic control flow graph for $T$ is given by $G = (V, E)$, where:
\begin{itemize}
\item
$V = \biguplus_{i=0}^n V_i$ is the disjoint union of the sets
of vertices $V_i$ of the individual phase CFGs $G_i$; and
\item
$E = (\biguplus_{i=0}^n E_i) \cup E_{\it dyn}$ is the disjoint union
of the sets of edges $E_i$ together with a set of {\em dynamic edges}
$E_{\it dyn}$ defined as follows:

$E_{\it dyn} = ({\it last}(\varphi_i), {\it first}(\varphi_{i+1}))$, where
${\it last}(\varphi_i)$ and ${\it first}(\varphi_{i+1})$ denote,
respectively, the basic blocks corresponding to the last instruction
of $\varphi_i$ and the first instruction of $\varphi_{i+1}$.
\end{itemize}
\end{definition}

\begin{figure}
\begin{minipage}{1.1in}
\begin{center}
\includegraphics[height=1.7in]{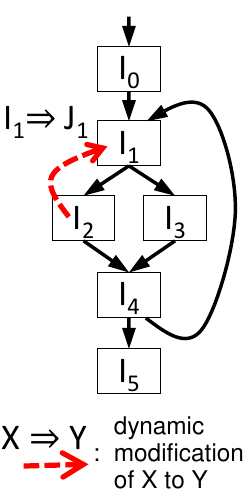}

$(a)$ Static CFG
\end{center}
\end{minipage}
\hspace{0.05in}
\begin{minipage}{2.1in}
\begin{center}
\includegraphics[height=1.7in]{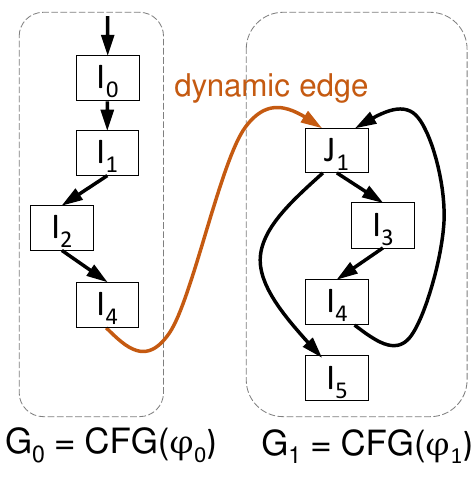}

$(b)$ Dynamic CFG
\end{center}
\end{minipage}

\caption{DCFG: An example}\label{fig:DCFG-example}
\end{figure}

\begin{example}\label{ex:dcfg}
Figure \ref{fig:DCFG-example} gives a simple example of a DCFG.
The static CFG of the program under consideration is shown in 
Figure \ref{fig:DCFG-example}$(a)$.
When instruction $I_2$ is executed, it changes instruction $I_1$ to $J_1$
(indicated by the dashed red arrow), where $J_1$ is a conditional branch
with possible successors $I_3$ and $I_5$. 
The following is an execution trace for this program along with its phases:
\begin{quote}
\begin{tabular}{lC{0.6in}C{0.7in}}
Trace: & \multicolumn{2}{c}{$\underbrace{\mbox{
                            \mbox{$I_0$}\ \ 
                            \mbox{$I_1$}\ \ 
                            \mbox{$I_2$}\ \  
                            \mbox{$I_4$}
}}
         \underbrace{\mbox{
	                  \mbox{$J_1$}\ \ 
			  \mbox{$I_3$}\ \ 
			  \mbox{$I_4$}\ \ 
			  \mbox{$J_1$}\ \ 
			  \mbox{$I_5$}}}$}\\
Phases & \ \ $\varphi_0$ & \multicolumn{1}{c}{$\varphi_1$}\\
\end{tabular}
\end{quote}
The first phase, $\varphi_0$, consists of the instruction sequence 
$I_0, I_1, I_2, I_4$.
When control returns to the top of the loop 
at the end of this sequence, instruction $I_1$ is found to have been changed
to $J_1$.  This ends $\varphi_0$ and begins
$\varphi_1$, which comprises the rest of the trace,
$J_1, I_3, I_4, J_1, I_5$.  The CFGs corresponding
to phases $\varphi_0$ and $\varphi_1$ in Figure \ref{fig:DCFG-example}$(b)$ are
$G_0$ and $G_1$ respectively.  Finally, the
control transfer from $\varphi_0$ to $\varphi_1$ is
indicated via a dynamic edge from the basic block of the last instruction
of $\varphi_0$ to the basic block of the first instruction in $\varphi_1$,
i.e., from the block for $I_4$ in $G_0$ to the block for $J_1$ in $G_1$.
\end{example}

The reader may notice, in Example \ref{ex:dcfg}, that the basic block containing
$I_4$ occurs in both $G_0$ and $G_1$.  This illustrates a potential drawback of
a naive implementation of DCFGs, namely, that CFG components may be replicated 
across different phases.  It is possible to implement DCFGs to avoid such
replication, but in this case it is important to ensure that
algorithms that traverse the DCFG (e.g., for slicing)
do not follow unrealizable paths.
The details are orthogonal to the primary thrust of this
paper and so are omitted; Section \ref{sec:performance} briefly sketches the
performance improvements we see from implementing sharing of DCFG components
across phases.

\subsubsection{Codegen Dependencies}
Dynamic code modification can induce a dependency between the code
performing the modification and the resulting modified code.  Consider
the following example:
\begin{center}
\begin{center}
\includegraphics[width=0.7\linewidth]{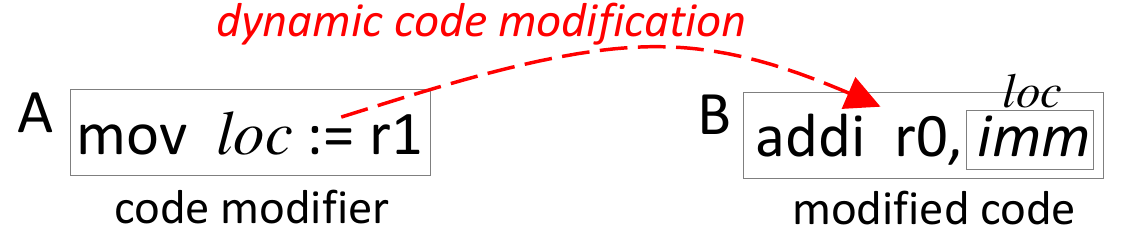}
\end{center}
\end{center}
In this example, \textsf{B} is an instruction that adds an immediate 
value \textit{imm} to the register \textsf{r0}; the bytes of \textsf{B} containing
\textit{imm} are at address \textit{loc}.
Thus, if \textit{loc} contains the value 5, then \textsf{B} is the instruction 
`\textsf{addi r0,5}'. Instruction \textsf{A} writes the
contents of register \textsf{r1} to address \textit{loc}, thereby
modifying \textsf{B}.  When \textsf{B} is executed,
the value added to \textsf{r0} depends on
the value written to
address \textit{loc} by \textsf{A}.  Thus, the execution of 
\textsf{A} affects the behavior of \textsf{B} through the act of
dynamic code modification, independent of any data or control dependencies
that may exist in the program.  We refer to dependencies arising in this way
due to dynamic code modification as {\em codegen dependencies}.  More formally:


\begin{definition}
Given an execution trace $T$ for a program, a dynamic instance of an
instruction $I \equiv T[i]$ is codegen-dependent on a dynamic instance of
an instruction $J \equiv T[j]$ $(j < i)$ if and only if, for some
${\it loc} \in \instr{I}$, the following hold:
\begin{enumerate}
\item
${\it loc} \in \writes{J}$, i.e., $J$ modifies the location {\it loc}\/; and
\item
$\forall k \mbox{ s.t. } j < k < i : {\it loc} \not\in \writes{T[k]}$, i.e., $J$ is the
instruction that most recently modifies {\it loc} before $I$ is executed.
\end{enumerate}
\end{definition}

While codegen dependencies resemble data dependencies in some
ways, it is different in one fundamental way.  If an instruction $I$ is data dependent
on an instruction $J$, then $J$ can change the values used by $I$, but not
the nature of the computation performed by $I$.  By contrast, if $I$ is codegen
dependent on $J$, then $J$ can change the nature of the computation performed
by $I$, e.g., from a {\em bitwise-or} instruction to a {\em jump-if-equal}
instruction as discussed earlier.

\subsection{DCFG Construction}\label{sec:pgm-repr}

Algorithm \ref{algo:dcfg} shows how we construct a DCFG from an execution trace.
The DCFG consists of a sequence of CFGs
$\{G_\varphi \mid \varphi = 0, 1, \ldots\}$, one per phase, linked together
by dynamic edges; we refer to the index $\varphi$ for these CFGs as their
phase index.  

Given an execution trace $T$, we proceed as follows.
We initialize the phase index $\varphi$ to 0 and the DCFG \textbf{G} to
$\emptyset$.  
The set $W$ of memory locations written in the current phase is 
initialized to $\emptyset$.
The CFG $G_\varphi$ is initialized to the empty graph and
added to \textbf{G} (line 7).  
We then iterate through the trace processing each instruction $T[i]$ in turn.
If $T[i]$ begins a new phase, we increment the phase index (line 10),
reset $W$ to $\emptyset$ (since no memory locations have been written in
the new phase that has just begun), initialize the CFG $V_\varphi$ for the
new phase to the empty graph, and add this new $V_\varphi$ to the DCFG \textbf{G}
(lines 10--12).
We then process the instruction $T[i]$ in the context of the CFG $G_\varphi$,
as discussed in Section \ref{sec:CFGs} (line 13).
At this point, if $T[i]$ is the first instruction of a phase (line 14),
it has been added to $G_\varphi$, which means $G_\varphi$ has a basic block
for it, so we add a dynamic edge from the basic block of the last instruction
of the previous phase to the basic block of the first instruction of the
current phase (line 15).  Finally, we update the set of written memory
locations by adding in the set of locations written by $T[i]$ (line 16).
We then continue the process with the next instruction of $T$.

\IncMargin{1em}
\begin{algorithm}  
 \DontPrintSemicolon
 \SetKwInput{KwData}{Input}
 \SetKwFunction{newphase}{newphase}
 \KwData{An execution trace $T$}
 \KwResult{A DCFG \textbf{G} for $T$}
 \BlankLine
    \SetKwFunction{FMain}{{\it instr\_starts\_new\_phase}}
    \SetKwProg{Fn}{function}{:}{}
    \Fn{\FMain{${\it Instr}, {\it WrittenLocs}$}}{
        \textbf{return} ($\instr{Instr} \cap {\it WrittenLocs} \neq \emptyset$)
 }
 \BlankLine
 \Begin {
    $\textbf{G} = \emptyset$\;
    $\varphi \longleftarrow \emptyset$\; 
    $W = \emptyset$\;
    $G_\varphi = (\emptyset, \emptyset)$; add $G_\varphi$ to \textbf{G}\;
    \For{$i = 0$ \KwTo ${\it len}(T)-1$} {
        \If {instr\_starts\_new\_phase(T[i], W)} {
            $\varphi \texttt{~+=~} 1$\;
            $W = \emptyset$\;
            $G_\varphi = (\emptyset, \emptyset)$; add $G_\varphi$ to \textbf{G}\;
        }
        process $T[i]$ in the context of $G_\varphi$ (see Sec. \ref{sec:CFGs})\;
        \If {instr\_starts\_new\_phase(T[i], W)} {
           add a dynamic edge from last block of $G_{\varphi-1}$ to first 
   	   block of $G_\varphi$\;
        }
        $W \longleftarrow W \cup \writes{T[i]}$\;
    }
}
\caption{DCFG Construction}\label{algo:dcfg}

\end{algorithm}

\subsection{Discussion}\label{sec:discussion}
The notions of phases, DCFGs, and codegen dependencies, as defined above, refer
to an execution trace, and thus to a particular execution of a program.  While
this suffices for the applications we were motivated by, it naturally raises the
question of how they might be extended to encompass multiple execution traces.

As a starting point, we note that for programs that do not have dynamic code
modification, every execution has a single phase, so the CFG constructed using
Algorithm \ref{algo:dcfg} corresponds to a path through its static CFG.  For such
programs, the CFGs obtained from multiple traces can be merged in a straightforward
way.

Dynamic code modification makes things more complex because different executions
of the same program can, in the fully general case, modify arbitrary parts of
the program and/or give rise to arbitrarily different dynamic code.
Specifying a single representation of that program that captures all of the
different possible code that can result from different executions seems challenging.
It may be possible to identify special cases that encompass real-life software
of interest, e.g., JIT compilers or dynamic binary rewriting systems, which do not
involve arbitrary and unrestricted kinds of dynamic code modification.
Since our motivating applications for this work did not warrant this additional
level of complexity, we leave such generalizations as future work.

\section{Applications}
\subsection{Program Slicing for Bug Localization and Exploit Analysis in JIT Compilers}\label{sec:slicing}
Program slicing refers to identifying instructions that (may) affect, or be affected
by, the value computed by an instruction in a program \cite{korel1997computation,%
  agrawal1990dynamic,Tip95}.  Slicing can be static or dynamic; and, orthogonally,
forward or backward.  By eliminating instructions that are provably irrelevant to
the computation of interest, slicing reduces the amount of code that has to be
examined in order to reason about it.

We implemented backward dynamic slicing as an application for evaluating
the efficacy of DCFGs and codegen dependencies for end-to-end reasoning about
dynamic code, with the goal of bug localization and exploit analysis in JIT compilers.  Backward dynamic
slicing aims to identify the set of instructions that may have affected the
value of a variable or location at some particular point in a particular execution
of the program.  Our implementation is based on Korel's algorithm for 
dynamic slicing of unstructured programs \cite{korel1997computation};
however, any slicing algorithm for unstructured programs would have been adequate. 

The algorithm, as originally defined, uses the usual notion of data dependencies
to find last definition of an instruction's operands: i.e., given an instruction
$I \equiv T[i]$ in an execution trace $T$, the {\em last definition} of $I$ is
the most recent instruction prior to position $i$ in $T$ such that $I$ is
data dependent on $J$, i.e., $J$ defines some source operand of $I$.  To work with
dynamic code, we modified this notion to also take codegen dependencies into
account, such that the last definition of an instruction $I \equiv T[i]$ in a
trace $T$ is the most recent instruction $J$ prior to $I$ in $T$ such that either
$(i)$ $I$ is data dependent on $J$; or $(ii)$ $I$ is codegen-dependent on $J$. 

It is important to note that the notion of DCFGs plays a crucial role in providing
control flow information needed to construct backward slices.
Analyses that reason about 
dynamic code solely through data dependencies, e.g. using taint propagation 
\cite{yin2007panorama,yin2010temu,henderson2014make,%
  chow2004understanding,enck2014taintdroid} are unable to capture the
effects of control dependencies and therefore are unsound with respect to slicing.

\subsection{Detecting Environmental Triggers in Malware}
Malware sometimes use environmental triggers to evade detection by performing
malicious actions only if the right environmental conditions are met, e.g.,
if the date has some specific value  or when certain keys are typed. 
Current work on detecting such behaviors is geared towards static
code, e.g., identifying conditional branches with input-tainted operands
\cite{brumley2008automatically}.  The idea is to use dynamic taint analysis
to identify conditional branches of the form
\begin{quote}
    \textbf{if} {\it expr} {\bf then} ${\it behavior}_1$ {\bf else} ${\it behavior}_2$
\end{quote}
where {\it expr} is tainted from (i.e., influenced by) some input values.  
Once such conditionals have been identified, it may be possible to use
other techniques, e.g., using SMT solvers to generate alternate inputs, to
further explore the program's behavior.

The presence of dynamic code potentially opens up other ways to implement
environmental triggers, e.g., 
by using the environmental input to directly affect the instruction bytes generated:
\begin{quote}
  trigger := {\it input}()\\
  codebuf := $f$(trigger)\\
  execute codebuf
\end{quote}
This idea can be illustrated by adapting an example of evasive behavior,
described by Brumley {\em et al.} \cite{brumley2008automatically}, to use dynamic code
instead of a straightforward conditional.
The code, shown in Figure \ref{fig:trigger},
uses bit-manipulation instead of conditionals to evaluate the trigger expression,
thereby rendering inapplicable techniques that rely on tainted conditionals.
The variable \texttt{day\_bits} is set to 1 or 0 depending on whether or not
the most significant bit of the value of the expression \texttt{day-9} is 0, i.e.,
whether or not the predicate \texttt{day $\geq$ 9} is true.  Similarly,
\texttt{mth\_bits} is 1 or 0 depending on whether or not 
\texttt{month $\geq$ 7} is true.  Thus, the 
variable \texttt{trigger} is 1 or 0 depending on whether the environmental 
trigger---in this example, the predicate
\texttt{day $\geq$ 9 \&\& month $\geq$ 7}---is true or not.  The assignment
to \texttt{*(addInstrPtr+11)} then writes this
value into the source byte of an assignment to a variable that is used in a
conditional to determine whether the malicious behavior is manifested.  
Note that the conditional that controls the execution of the \texttt{payload()} function
is neither data-dependent nor control-dependent on the input; instead there is a
codegen dependency between this conditional and the patching instructions, 
which are data dependent on the input.

\begin{figure}
  {\footnotesize
\begin{verbatim}
void hide() {
    volatile int environmental_trigger = 0;
    if (environmental_trigger) {
        payload(...);    // perform malicious action
    }
}

void patch() {
    int pg_sz = sysconf(_SC_PAGE_SIZE); 
    mprotect((void*) ((((long) &hide) / pg_sz) * pg_sz), 
             pg_sz * 2, PROT_READ | PROT_WRITE | PROT_EXEC);
    time_t rawtime;
    struct tm * systime;
    time(&rawtime);
    systime = localtime(&rawtime);

    int day = systime->tm_mday;
    int day_test = ~(day - 9);
    int day_bits = day_test >> 31;  // day_bits == 1 iff day >= 9

    int month = systime->tm_mon+1;
    int mth_test = ~(month - 7);
    int mth_bits = mth_test >> 31;  // mth_bits == 1 iff month >= 7

    // trigger == 1 iff (day >= 9 && month >= 7)
    int trigger = day_bits & mth_bits; 

    unsigned char* addInstrPtr = ((unsigned char*) &hide);
    *(addInstrPtr+11) = trigger;
}

int main() {
    hide();
    patch();
    hide();
    return 0;
}
\end{verbatim}
}

\caption{Environmental trigger based on dynamic code}\label{fig:trigger}
\end{figure}

\begin{figure*}
\begin{minipage}{\columnwidth}
{\footnotesize\tt
\begin{tabbing}
XXX\= XXXXXXXXXXXXXXXXXXXX\=\kill
main:\\
\>    	pushq   \%rbp\\
\>    	movq	\%rsp, \%rbp\\
\>    	xor 	\%rax, \%rax\\
L0:\\
\>    	leaq	L2(\%rip), \%rcx\\
\>    	leaq	L0(\%rip), \%rdx\\
\>    	sub 	\%rdx, \%rcx\\
\>    	imul	\%rax, \%rcx\\
\>    	leaq	L1(\%rip), \%rdx\\
\>    	add 	\%cl, 1(\%rdx)\> \#\#\# \textbf{<<< Gen1}\\
\>    	inc 	\%rax\\
L1:\\
\>    	jno 	L0 \> \#\#\# \textbf{<<< Dyn1}\\
L2:\\
\>    	nop\\
\>    	movq \$60, \%rax\\
\>    	movq \$0,  \%rdi\\
\>    	syscall\\
\end{tabbing}

}
\begin{center}
  $(a)$ Synthetic Benchmark 1
\end{center}
\end{minipage}
\hspace{0.1in}
\begin{minipage}{\columnwidth}
{\footnotesize\tt
\begin{tabbing}
XXXX\=XXXX\=XXXX\=XX\=XX\=XX\=XXXXXXXXXXXXXXXXX\=\kill
// Source: https://www.quora.com/C-programming-language-Can-\\
// you-write-a-C-program-to-demonstrate-a-self-modifying-code\\
\\
int AddTwo(int input) \{\\
\>    return input + 2;  \>\>\>\>\>\>// \textbf{<<< Dyn2}\\
\}\\
\\
void patchFunction() \{\\
\>    int pageSize = sysconf(\_SC\_PAGE\_SIZE); \\
\>    mprotect((void*) ((((long) \&AddTwo) / pageSize) * pageSize), \\
\>\>\>\>\> pageSize * 2, PROT\_READ | PROT\_WRITE | PROT\_EXEC);\\
\\
\>    // Change the byte used in addition in AddTwo()\\
\>    unsigned char* addInstrPtr = ((unsigned char*) \&AddTwo);\\
\>    *(addInstrPtr+12) = 3;  \>\>\>\>\>\>// \textbf{<<< Gen2}\\
\}\\
\\
int main() \{\\
\>    int out = AddTwo(2);\\
\>    patchFunction();\\
\>    AddTwo(2);\\
\>    return 0;\\
\}
\end{tabbing}
}
\begin{center}
  $(b)$ Synthetic Benchmark 2
\end{center}
\end{minipage}
\caption{Source code for Synthetic Benchmarks}\label{fig:synthetic}
\end{figure*}

\section{Evaluation}\label{sec:eval}

\begin{table}
  {\small
  \begin{tabular}{|L{0.6cm}|L{2.0cm}|C{0.9cm}|C{0.8cm}|C{0.8cm}|C{0.8cm}|}
     \cline{3-6}
     \multicolumn{2}{c|}{~}
     & Our~approach & PinPlay & angr & Triton\\
     \hline
     \multirow{2}{*}{\STAB{
         \begin{minipage}{1.5cm}{\it\small Synth-\\etic}\end{minipage}}}
        & Benchmark 1 & Y & N & N & N\\
        & Benchmark 2 & Y & N & N & N\\
    \hline
    \hline
     \multirow{3}{*}{\STAB{\rotatebox{90}{
         \begin{minipage}{1.5cm}{\it\small Exploit\\analysis}\end{minipage}}}}
        & V8 OOB to JIT & Y & X & X & X\\
        & code pages    &   &   &   &  \\
        \cline{2-6}
        & V8 Escape analysis bug & Y & X & X & X\\ \cline{2-6}
        & LuaJIT exploit & Y & N & N & N\\ \hline\hline
     \multirow{2}{*}{\STAB{\rotatebox{90}{
         \begin{minipage}{1.5cm}{\it\small Bug\\localization}\end{minipage}}}}
        & OOB Read & Y & X & X & X\\
        \cline{2-6}
        & JIT type confusion & Y & X & X & X\\
        \cline{2-6}
        & Scoping issue & Y & X & X & X\\
    \hline
  \end{tabular}
  }
\ \\
\medskip

\begin{tabular}{R{0.5cm}l}
{\bf Key:} & \\
Y: & Picks up dynamic code generator from backwards slice\\
   & of dynamic code.\\
N: & Does not pick up dynamic code generator from backwards\\
   & slice of dynamic code.\\
X: & Crashes or fails to load.\\
\end{tabular}
\ \\
\ \\
  \caption{Slicing: Soundness}\label{tab:precision}
\end{table}

\subsection{Overview}
We built a prototype implementation to evaluate the efficacy of our ideas.
We used Intel's Pin software (version 3.7) \cite{pin} for program
instrumentation and collecting instruction-level execution traces; and
XED (version 8.20.0) \cite{xed} for instruction decoding.
We iterate over the instruction trace to construct a DCFG for the execution.
We identify dynamic code as follows.
We use our own tainting library during construction of our DCFGs to 
determine code-gen dependencies: we taint writes to memory, with each
memory write getting a distinct taint label.  For each 
instruction in the trace we check whether any of its instruction bytes
is tainted, in which case the instruction is flagged as dynamic.

 We ran our experiments on a machine with 32 cores (@ 3.30 Ghz) and 1 terabyte of RAM, running 
 Ubuntu 16.04. 
 We used both synthetic and real-world examples for our evaluation, and
compared our results with three widely-used modern dynamic analysis tools that
 implement backwards slicing, namely: PinPlay \cite{patil2010pinplay} (revision 1.29), 
 angr \cite{shoshitaishvili2016state,stephens2016driller} (commit bd3c6d8 on github), and 
 Triton \cite{SSTIC2015-Saudel-Salwan} (build number 1397),
 to see whether the slices they computed captured the buggy JIT-compiler code.
 For these experiments, we invoked these tools to incorporate
 support for self-modifying code as follows:
 we set the flags \texttt{smc\_support} and \texttt{smc\_strict} flags to true for
 PinPlay, and loaded our project with \texttt{auto\_load\_libs} and
 \texttt{support\_selfmodifying\_code} set to true for angr.

\subsection{Slicing}

\subsubsection{Soundness}
A sound slicing algorithm does not exclude any statement that can influence the slicing
criterion.  In our case, this means that
given some observed behavior
in dynamic code that is the result of some logic in the code that performs the
dynamic code generation, a sound slicing algorithm should be able to identify the connection
between the observed behavior in the dynamically generated code and the culprit
code in the dynamic code generator.

\paragraph{5.2.1.1. Synthetic Benchmarks}
We used two small examples of dynamic code as synthetic benchmarks.
These  implement relatively small and simple cases of dynamic code and allow us to
focus on the core concepts underlying our approach.

The first example, shown in Figure \ref{fig:synthetic}$(a)$, is a small program 
written in x86 assembly code.\footnote{This code is compiled using the command
  `\texttt{gcc --static -g -Wl,--omagic pgm.s -o pgm.out}'; the
  `\texttt{--omagic}' option passed to the linker makes the text section writable
  and makes it unnecessary to do so explicitly from within the code.}
The instruction tagged as `\texttt{Gen1}' in Figure \ref{fig:synthetic}$(a)$
performs dynamic code modification by adding the contents of register
\texttt{\%cl} to the memory location with address given by \texttt{1(\%rdx)}.
The operand of the instruction tagged `\texttt{Dyn1}' is modified as a result.
The instructions \texttt{Gen1} and \texttt{Dyn1} both occur in a loop, with
\texttt{Dyn1} being the backward conditional jump at the end of each iteration
of the loop.
When this program is executed, \texttt{Gen1} repeatedly dynamically modifies the
offset for the backward jump of the loop, so that the target of the jump changes with
each iteration, until eventually it turns into a forward jump whose target is
outside the loop.  An examination of the execution trace of this program
shows that \texttt{Gen1} writes to the second byte of \texttt{Dyn1}, inducing
a codegen dependency from \texttt{Dyn1} to \texttt{Gen1}.  
Each time we encounter \texttt{Dyn1} after it 
has been modified in this way we get a new phase.

The second example, shown in Figure \ref{fig:synthetic}$(b)$, is
taken from an online discussion forum
\cite{bhoi2018selfmod}.  It has a function \texttt{AddTwo()} that returns the value
obtained by adding 2 to its integer argument.  The instruction that performs
this addition, tagged as `\texttt{Dyn2}' in Figure \ref{fig:synthetic}$(b)$, is
modified at runtime to change the value added to the argument from 2 to 3.
The instruction that performs this dynamic code modification is tagged
`\texttt{Gen2}'.
Thus, \texttt{Dyn2} is codegen dependent on \texttt{Gen2}.  This example has two
phases, with a new phase being generated upon encountering the new version of
\texttt{Dyn2} after it is patched.

We used our backwards slicing implementation, discussed in Section \ref{sec:slicing},
on these examples for an initial evaluation of our program representation. 
In each case, we computed a backward dynamic slice from the dynamic code:
namely, the last instance of the instructions labeled \texttt{Dyn1} and 
\texttt{Dyn2} as discussed above. 
In both cases the slice criteria included the 
dynamic instruction modified for both examples, respectively.
We found that, in both cases, the slices so computed contained the instructions
that modify/write dynamic code.  Codegen dependencies and DCFGs play a crucial
role in this: for example, 
the instruction \texttt{Gen2} in the second example is explicitly picked up  due to
a codegen dependency from the patched instruction \texttt{Dyn2}.  There are
no data or control dependencies between the instructions 
Gen2 and Dyn2, so traditional dependencies alone would not have been sufficient
to pick up the code that does the dynamic code modification.

To assess the current state of the art in analysis of dynamic code,
we used three popular dynamic analysis tools: PinPlay, angr, and Triton,
to compute backward dynamic slices on the results computed by these programs.
These results are summarized in Table {tab:precision}.
Our experiments indicate that (1) the CFGs constructed by these tools
do not represent the different versions of code resulting from dynamic
code modification; and (2) while all three tools successfully included all of the 
relevant non-codegen-dependent instructions in the slices they computed, 
none of them are able to pick up the code that performs dynamic modification.

\paragraph{5.2.1.2. Real World Examples:}
To evaluate our approach on real world software that uses dynamic code, we
consider three example applications:
(1) analysis of exploits involving JIT code;
(2) bug localization in JIT compilers; and
(3) detection of trigger-based evasive behaviors that use dynamic code.
Our goal was to perform end-to-end analyses on these examples, i.e.,
start from the problematic dynamic code and compute a backward dynamic slice that
includes portions of the dynamic code generator where the bug/security 
exploit originates.  The results
are shown in Table \ref{tab:precision}.

\begin{table*}
  {\small
  \renewcommand{\arraystretch}{1.3}
  \begin{tabular}{
        |L{0.5cm}
        |L{3.0cm}
  	|R{1.5cm}|R{0.7cm}					
	|R{0.9cm}|R{0.9cm}|R{0.9cm}|C{0.9cm}|R{0.9cm}		
	|R{0.9cm}|R{1.1cm}|R{0.8cm}|				
  }
  \cline{3-12}
     \multicolumn{2}{c|}{~}
     & \multicolumn{2}{|c|}{\sc Tracing}
     & \multicolumn{5}{|c|}{\sc DCFG Construction}
     & \multicolumn{3}{|c|}{\sc Slicing}
     \\ \cline{2-12}
        \multicolumn{1}{c|}{~}
        & \textit{Test program}
	& \multicolumn{1}{|c|}{$N_{\it trace}$}
	& \multicolumn{1}{|c|}{$T_{\it read}$}
	& \multicolumn{1}{|c|}{$N_{\it instrs}$}
	& \multicolumn{1}{|c|}{$N_{\it blocks}$}
	& \multicolumn{1}{|c|}{$N_{\it edges}$}
	& \multicolumn{1}{|c|}{$N_{\it phases}$}
	& \multicolumn{1}{|c|}{$T_{\it DCFG}$}
	& \multicolumn{1}{|c|}{$N_{\it slice}$}
	& \multicolumn{1}{|c|}{$T_{\it slice}$}
	& \multicolumn{1}{|c|}{$\Delta_{\it slice}$}
	\\ \hline
     \multirow{3}{*}{\STAB{\rotatebox[origin=c]{90}{
         \begin{minipage}{1.0cm}{\it\small Exploit\\analysis}\end{minipage}}}}
     & V8 OOB to JIT Code Pages 
     & 11,134,237 & 10.68 
     & 191,613 & 41,302 & 117,158 & 4 & 146.88 
     & 81,986 & 433.25 & 57 \% 
     \\ \cline{2-12}
     & V8 Escape analysis bug 
     & 135,295,168 & 130.76
     & 245,935 & 52,929 & 153,922 & 3 & 1,793.23 
     & 120,885 & 10,193.08 & 50 \% 
     \\ \cline{2-12}
     & LuaJIT Exploit 
     & 464,743 & 0.60 
     & 18,248 & 4584 & 12,606 & 2 & 7.47 
     & 5,139 & 7.76 & 71 \% 
     \\ \hline\hline
     \multirow{2}{*}{\STAB{\rotatebox[origin=c]{90}{
         \begin{minipage}{1.5cm}{\it\small Bug\\localization}\end{minipage}}}}
     & OOB Read 
     & 14,720,437 & 14.25 
     & 150,115 & 31,469 & 92,254 & 2 & 196.29 
     & 61,511 & 579.78 & 59 \% 
     \\ \cline{2-12}
     & JIT Type Confusion 
     & 9,663,365 & 9.49 
     & 158,849 & 32,536 & 93,132 & 9 & 130.26 
     & 67,765 & 146.47 & 57 \% 
     \\ \cline{2-12}
     & Scoping issue 
     & 7,882,295 & 7.56 
     & 99,378 & 22,394 & 62,204 & 4 & 102.31 
     & 47,023 & 970.95 & 52 \% 
     \\ \hline
  \end{tabular}
  }
  \renewcommand{\arraystretch}{1.0}
\ \\
\medskip
\ \\
\medskip
\begin{tabular}{ll}
{\bf Key:} & \\
  \begin{tabular}{lcl}
  $N_{\it trace}$ & : & No. of instructions in execution trace\\
  $T_{\it read}$ & : & Time to read trace (seconds)\\
  $N_{\it instrs}$  & : & No. of instructions in DCFG\\
  $N_{\it blocks}$  & : & No. of basic blocks in DCFG\\
  $N_{\it edges}$  & : & No. of basic blocks in DCFG\\
  \end{tabular}
  &
  \begin{tabular}{lcl}
  $N_{\it phases}$ & : & No. of phases\\
  $T_{\it DCFG}$ & : & DCFG construction time (seconds)\\
  ${\it N}_{\it slice}$ & : & No. of instructions in slice\\
  $T_{\it slice}$ & : & Slice construction time (seconds)\\
  $\Delta_{\it slice}$ & : & Fraction of DCFG removed from slice \\
 	& & = $(N_{\it instrs}-{\it N}_{\it slice})/N_{\it instrs}$.
  \end{tabular}
\end{tabular}
\ \\
\ \\
  \caption{Slicing: Performance}\label{tab:slicing-performance}
\end{table*}


\paragraph{\bf a. Exploit Analysis:}
We consider three examples of exploits involving dynamic code:
\begin{enumerate}
\item
malicious shellcode originating from an out-of-bounds
(OOB) write to the JIT code pages in
Google's V8 JavaScript engine \cite{v8oobexploit};
\item
escape analysis bug in V8's JIT compiler (CVE-2017-5121)
\cite{rabet2017}; and
\item
malicious bytecode used to escape a LuaJIT sandbox \cite{luajitexploit}.
\end{enumerate}
We analyzed each of these exploits as follows.  We used the proof-of-concept
code to compute a DCFG/backward dynamic slice starting from the dynamically generated
exploit code. Separately, we examined the write ups for each of the exploits to determine the bugs responsible for each exploit, identifying the buggy code generator portions in the execution traces recorded for each exploit. We then
checked the slice to determine whether the buggy generator code is present in the slice.

 The first security exploit we consider entails an OOB write 
to the JIT code pages within Google's V8 JavaScript engine \cite{v8oobexploit}. The exploit is a result of array type ambiguity, which allows for the author to construct a non-linear overflow with which to write and execute arbitrary shell code. The original author demonstrated this by launching xcalc. We modified the 
same exploit so that the shellcode instead encounters a divide-by-zero exception to aide in identifying the dynamic code in the execution trace.
We used D8 to invoke the V8 JavaScript engine from the command line,
obtained an execution trace using Pin, and then constructed a DCFG along with a backward dynamic
slice from the first nop shellcode instruction that appears in the nop sled
immediately before the divide-by-zero exception.
Our backward slice successfully included both the buggy code within V8 that led to the array type ambiguity along with the 
instructions that actually generated the shellcode at runtime.

The second exploit we examined is discussed in detail by Rabet \cite{rabet2017}.
It arises out of a bug in V8's escape analysis and causes some variable
initializations in the JIT-optimized code to be incorrectly optimized away when performing load reduction.
The proof-of-concept code provided causes V8 to crash while executing the optimized dynamic code due to an OOB read. The write up provided by Rabet proceeds to use this OOB read as a stepping stone towards  demonstrating arbitrary code execution. For our analysis of this example, we built our DCFG from the execution trace recorded by Pin and then we computed a backward
dynamic slice from the dynamic instruction prior to the exception that is thrown due to the OOB read.  We found that the resulting
slice correctly included the buggy portions of the load reducer in the escape 
analysis phase of V8's JIT compiler, whose optimizations cause the OOB read.

Our final example in this category was with malicious Lua bytecode being used 
to escape a sandbox in LuaJIT \cite{luajitexploit}. The proof of concept malicious program corrupts bytecode with the goal of writing shellcode which prints a message.
We followed an approach similar to the one we used to slice the V8 OOB write, starting our slice at the beginning of the 
NOP sled used in the attack.  We found that the backward slice computed by our
tool correctly picks up the Lua code that generates the shellcode.

In summary, using our approach we are able to compute backward slices starting
at some problematic dynamically created instruction(s) and working back to
identify the JIT compiler logic that created that instruction. Our slices make use of our DCFGs when determining removable instructions, allowing for multiple versions of dynamic instructions to be reasoned about separately. Furthermore, the addition of code-gen dependencies from our DCFGs allow for our backward slices to reason about dependencies between the code that does the dynamic modification and the dynamically generated code itself.
This can be the starting point for further analysis, e.g.,
the slice can be used to determine where and how the shellcode for the exploit
is crafted, whether or not the bytecode was corrupted during execution, etc.
Such explorations are facilitated by the ability to obtain program slices that
contain $(a)$ all versions of the dynamic code executed; and $(b)$ the code
that generates the dynamic code, linked explicitly through codegen 
dependencies.

To better understand how well current approaches apply to such analysis problems,
we also used these examples to experiment with PinPlay, angr, and Triton.  The
results are summarized in Table \ref{tab:precision}.

\paragraph{\bf b. Bug Localization:}
We consider three JIT compiler bugs from Google's V8 JavaScript engine that were
posted to \texttt{bugs.chromium.org} and classified as ``Type: Bug-Security.''
\begin{enumerate}
\item
  Empty jump tables generated by the bytecode generator leading to
  out-of-bound reads that cause the generated JIT-compiled to crash 
  (Issue 794825) \cite{v8bug1}.

\item
A type confusion bug that leads to a crash after the dynamic code has
been generated (Issue 794822) \cite{v8bug2}.

\item
  Arrow function scope fixing bug, where the system cannot handle
  certain constructs involving a single line arrow function,
  resulting in a crash (Issue 807096)
  \cite{v8bug3}.

\end{enumerate}
For each of these bugs we proceeded as follows.  
To identify the problematic code in the JIT compiler, we examined the
corresponding GitHub commits, together with any relevant information in the
bug report, to determine the code that was changed to fix
the bug.  We delineated the problem code so identified using small
``marker code snippets''---i.e., small easily identifiable code snippets
that do not affect the operation of the JIT compiler---and confirmed
that the behavior of the buggy JIT compiler was unaffected. 
We then used the example code submitted with the bug report to obtain an 
execution trace demonstrating the bug, and used this trace, together with the
DCFG constructed from it, to compute a backward dynamic slice starting
from the instruction that crashed.  Finally, we analyzed the resulting slice
to determine whether the problematic code, as identified above, was included 
in the slice.

The results of our experiments are summarized in Table \ref{tab:precision}.
Our end-to-end analysis was able to successfully pick up the buggy 
code for each of the bugs mentioned above in the slice, allowing one to narrow 
down the functions involved in V8 that lead to the crash.


To assess the efficacy of current state-of-the-art program
analysis tools on these examples, we experimented with same dynamic
analysis toolkits considered earlier, namely, PinPlay, angr, and Triton.
None of them were able to successfully analyze the examples
involving V8: they all crashed with internal errors when loading V8.
All three tools were able to process the LuaJIT example without crashing,
but none of the slices they computed contained the JIT-compiler or exploit code that
created the dynamic code.

\subsubsection{Performance}\label{sec:performance}
Table \ref{tab:slicing-performance} shows the performance of our prototype
DCFG-based slicing implementation on our real-world test inputs (the environmental
trigger example is omitted because it does not use backward slicing).  These input
programs all involve computations of substantial size: the smallest, LuaJIT
exploit, has a trace of 464K instructions, while the remaining execution traces
range from almost 7.9M instructions (V8 scoping issue bug)
to 135M instructions (V8 escape analysis bug).  The time taken to read the
traces (and do nothing else) is roughly 1M instructions/sec.\footnote{Our
  implementation uses Pin to collect an instruction trace that is written to
  a file on disk.  The numbers reported here refer to the time required to read
  such instruction trace files; the time taken to record the traces and write the
  trace files, which depends on the tracing tool used and is independent of the
  ideas described here, is not included.}

The DCFGs constructed typically range in size from about 22K basic blocks and 62K
edges (V8 scoping issue bug) to about 41K blocks and 117K edges (V8 OOB exploit),
with a low of 4.6K blocks and 12K edges for the LuaJIT exploit and a high of
about 53K blocks and 154K edges for the V8 escape analysis bug.  Most of our test
programs have $2-4$ phases, with the V8 JIT type confusion example an
outlier with 9 phases.  DCFG construction incurs an overhead of roughly
$15\times$ over simply reading a trace: most of the test inputs take roughly
$2-3$ minutes, with the lowest time being 7.5 seconds for the LuaJIT exploit and
the highest being about 30 minutes for the V8 escape analysis bug.  Since DCFG
construction involves processing each instruction in the execution
trace, the time taken depends on the sizes of both the instruction trace
and the DCFG.

The overhead incurred by slicing relative to the time taken for DCFG construction
ranges from $1.04\times$ for the LuaJIT exploit to $9.5\times$ for the V8 scoping
issue bug, with most of the test programs ranging from $3\times$ to $6\times$.
In absolute terms, most of the programs take about $2-10$ minutes for slicing,
with a low of about 8 secs for the LuaJIT example and a high of about about 2.8
hours for the V8 escape analysis bug.
Slicing is able to remove about 50\%--60\% of the instructions in the DCFG, with
a high of 71\% of the instructions removed for the LuaJIT exploit.  These
results are further improved by excluding code that is irrelevant to the JIT-compiler,
i.e., the front-end parser, byte-code generator, and interpreter.  These numbers
indicate that our approach is both practical (in terms of time) and useful
(in terms of the amount of code removed from the DCFG).  Since our approach
does not fundamentally alter the slicing algorithm, but rather augments it to
work over DCFGs and use codegen dependencies, it is not difficult to adapt our
approach to other slicing algorithms with different cost-precision characteristics.

Finally, an important aspect of usability is that of scalability.  This is
especially crucial for dynamic analyses, since the complexity of real-world
software that use dynamic code, combined with the fact that they are often
multi-threaded, means that instruction traces can become quite large.  In this
context, we note that our prototype implementation, which was built without any
particular attention to performance, is nevertheless resonably
scalable and is able to handle computations of considerable size on software of
real-world complexity (e.g., a trace of 135M instructions from a widely-used
JavaScript engine, corresponding to a CFG containing some 53K basic blocks and
245K instructions).

As noted in Section \ref{sec:DCFG}, a naive implementation of DCFGs can
lead to DCFG components being replicated across multiple phases, resulting in
a lot of of wasted memory.  Our implementation of DCFGs avoids 
such replication by associating a set of phase numbers with each instruction,
basic block, and edge in the DCFG; our slicing algorithm is modified to
avoid following unrealizable paths in the DCFG.  
For the real-world benchmarks we tested, this results in a savings
of about 18\% in DCFG size (no. of basic blocks).

\begin{table*}
  {\small
  \renewcommand{\arraystretch}{1.3}
  \begin{tabular}{
        |L{0.5cm}
        |L{3.8cm}
        |c
	|c
	|c
	|c
	|c
	|c
	|c
	|
  }
  \cline{3-9}
     \multicolumn{2}{c|}{~}
     & \multicolumn{2}{|c|}{\sc Original}
     & \multicolumn{2}{|c|}{\sc Dicing}
     & \multicolumn{3}{|c|}{\sc Improvement (\%)}
     \\ \cline{2-9}
        \multicolumn{1}{c|}{~}
        & \textit{Test program}
	& \multicolumn{1}{|c|}{${\it DCFG}_{\it orig}$}
	& \multicolumn{1}{|c|}{${\it slice}_{\it orig}$}
	& \multicolumn{1}{|c|}{${\it DCFG}_{\it mk}$}
	& \multicolumn{1}{|c|}{${\it slice}_{\it mk}$}
	& \multicolumn{1}{|c|}{$\Delta_{\it DCFG}$}
		& \multicolumn{1}{|c|}{$\Delta_{\it slice}$}
	& \multicolumn{1}{|c|}{$\Delta_{\it mk}$}
     \\ \hline
     \multirow{3}{*}{\STAB{\rotatebox[origin=c]{90}{
         \begin{minipage}{1.0cm}{\it\small Exploit\\analysis}\end{minipage}}}}
     & V8 OOB to JIT Code Pages 
     & 191,613 
     & 81,986
     & 90,736
     & 42,317
     & 52.6 & 48.4 & 53.4 
     \\
     \cline{2-9}
     & V8 Escape analysis bug 
     & 245,935 
     & 120,885
     & 157,847
     & 89,307
     & 35.8 & 26.1 & 43.4
     \\
    \cline{2-9}
     & LuaJIT Exploit 
     & 18,248 
     & 5,139 
     & 10,354 & 1,808 & 43.2 & 64.8 & 82.5\\
     \hline\hline
     \multirow{2}{*}{\STAB{\rotatebox[origin=c]{90}{
         \begin{minipage}{1.5cm}{\it\small Bug\\localization}\end{minipage}}}}
     & OOB Read 
     & 150,115 
     & 61,511
     & 35,261
     & 10,460
     & 59.0 & 83.0 & 70.3
     \\
     \cline{2-9}
     & JIT Type Confusion 
     & 158,849 
     & 67,765
     & 188
     & 103 
     & 99.9 & 99.8 & 45.2
     \\
     \cline{2-9}
     & Scoping issue 
     & 99,378
     & 47,023
     & 14,896
     & 7,721
     & 85.0 & 83.6 & 48.2
     \\
     \hline
  \end{tabular}
  }
  \renewcommand{\arraystretch}{1.0}
\ \\
\medskip
\begin{tabular}{l}
{\bf Key:} \\
  \begin{tabular}{lcl c lcl}
  ${\it DCFG}_{\it orig}$ & : & No. of instructions in original DCFG
  & & $\Delta_{\it DCFG}$ & : & Improvement in DCFG size due to dicing\\
  
  ${\it slice}_{\it orig}$ & : & No. of DCFG instructions in original slice
  & & & & = $({\it DCFG}_{\it orig} - {\it DCFG}_{\it mk})/{\it DCFG}_{\it orig}$\\
  
  ${\it DCFG}_{\it mk}$  & : & No. of instructions in DCFG with marker
  & & $\Delta_{\it slice}$ & : & Improvement in slice size due to dicing\\
  
  ${\it slice}_{\it mk}$  & : & No. of DCFG instructions in slice with marker
  & & & & = $({\it slice}_{\it orig} - {\it slice}_{\it mk})/{\it slice}_{\it orig}$\\
  
  & & & & $\Delta_{\it mk}$ & : & Fraction of ${\it DCFG}_{\it mk}$ removed due to dicing\\
  & & & & & & = $({\it DCFG}_{\it mk} - {\it slice}_{\it mk})/{\it DCFG}_{\it mk}$\\
  \end{tabular}
\end{tabular}
\ \\
\ \\
  \caption{Dicing: Performance}\label{tab:dicing-performance}
\end{table*}


\subsubsection{Focusing the analysis: markers and dicing}\label{markers}
Given the overall objective of localizing problems in the culprit JIT-compiler
code, it is useful to examine the extent to which our approach
is able to reduce the amount of actual JIT-compiler code that has to be considered:
for example, a slice that excludes much of the front-end parser and interpreter,
but includes the entirety of the JIT compiler, would not be very useful.
To evaluate this, we repeated our experiments to focus the our evaluation,
as much as possible, on the JIT-compiler portion of the system.  To do this, we
placed markers---i.e., small code snippets that are unambiguously identifiable
and semantically neutral---in the code as close as we were able to the
invocation of the JIT compiler.  During analysis, we excluded the portion
of the execution trace before the marker.
This effectively computed a program dice that excluded the front-end parser, 
byte-code generator, and interpreter.

Table \ref{tab:dicing-performance} gives the results of these experiments.
The two columns labeled `{\sc Original}' refer to the size of the DCFG and the
backward slice computed without markers, i.e., as shown in Table \ref{tab:slicing-performance};
the columns labeled `{\sc Dicing}' refer to the size of the DCFG and slice when
markers are used; the columns labeled `{\sc Improvement}' show the percentage
improvement due to dicing.  The columns labeled $\Delta_{\it DCFG}$ and
$\Delta_{\it slice}$ show, respectively, the reductions in the size of the DCFG
and the slice when irrelevant code is excluded.  These are in the range
35\%--85\% for DCFG size and 26\%--84\% for slice size.
The JIT Type Confusion bug sample is an outlier, with almost all of the original
DCFG and slice eliminated.
The final column, labeled $\Delta_{\it mk}$, shows the effects of slicing focusing
only on the DCFG resulting from dicing: these range from about 43\% to about 82\%.
Overall, these results show that (1) our approach is effective in focusing on the relevant portions
of the JIT compiler; and (2) the use of code markers to identify entry into
the JIT compiler can be helpful in zeroing in on the relevant portions of the code
being analyzed.

\subsection{Detecting Environmental Triggers}

Our implementation to detect environment-dependent dynamic code relies on the fact
that there must be a codegen dependency from code that is data or control
dependent on some environmental value to some code that is dynamically generated.
We first collect an execution 
trace and build our program representation. Next, we taint the input source and 
propagate the taint forward.  If, at any point, we find 
a codegen dependency from an instruction with tainted operands to an
instruction that is later executed, it means that an input-dependent value
may be influencing the instruction bytes of some dynamic instruction.  In
this case, we report that 
there is dynamic input-dependent program behavior. 

\begin{figure}
  {\footnotesize
\begin{verbatim}

void patch() {
    ...
    int final = day_bits & mth_bits; 
    // Implicit flow here
    volatile int x = 0, y = 0, z = 0;
    if(final){
      x = 1;
    } else {
      y = 1;
    }
    if(x == 0){
      z = 0;
    } 
    if(y == 0){
      z = 1;
    }
    unsigned char* addInstrPtr = ((unsigned char*) &hide);
    *(addInstrPtr+11) = z;
}
\end{verbatim}
}

\caption{Environmental trigger based on dynamic code with implicit flow}\label{fig:trigger2}
\end{figure}

As mentioned in Section \ref{sec:slicing}, data dependencies alone may not suffice
for reasoning about information flow in dynamic code. We demonstrate 
this using a variant of the previous example but
uses an implicit flow to propagate the value of the
environmental trigger; the implicit flow code,
shown in Figure \ref{fig:trigger2}, is based
on an example from Cavallaro {\em et al.} \cite{cavallaro2008limits}.
To analyze this example, we compute a backward dynamic slice with the slicing
criterion being the dynamically modified code location at the point where it
is executed.  This slice correctly includes the environmental triggers,
indicating that the code modification is influenced by environmental triggers.
The slice also includes the relevant portion of the implicit flow code, therey
demonstrating the utility of the control flow information obtained from the
DCFG.
In practice, one might use forward slicing from the point where
triggers are evaluated to determine whether any dynamically generated code
is in the slice.  However, our point here is not so much one
of the directionality of the computed slice, but rather to demonstrate the
sophistication of analyses possible when reasoning about dynamic code
using control flow information, available in DCFGs, together with codegen dependencies.

Our implementation correctly detects that environmental values influence
dynamic program behavior for the examples shown in Figures \ref{fig:trigger} and \ref{fig:trigger2}.
When tested with S2E, a widely used 
symbolic execution engine, we found that the input values used to patch the
function \texttt{hide()} are silently concretized and only the false path is
explored. The popular analysis tool angr gave a similar result. 


\subsection{The role of codegen dependencies}
Program slices are not guaranteed, in general, to be minimal. This means that 
it is possible, in theory, for an instruction in the slice to be included due 
to imprecision in the slicing algorithm. To determine whether imprecision may
have played a role in the results reported above, we re-ran our slicing tool 
on each of the examples discussed above with a directive to ignore codegen dependencies.

For the exploit examples discussed above, our slices were computed
starting at a \texttt{NOP} instruction in the NOP sled generated as part of
the shellcode.  For each example, ignoring codegen dependencies
resulted in the computation of a slice consisting of just the \texttt{NOP} instruction
itself and nothing else.  By contrast, when codegen dependencies were included
as part of the slice computation, the relevant JIT-compiler code was included
in the slice.  This demonstrates that codegen dependencies play a crucial role
in the computation of the slice linking the shellcode to the code that generated it.

For the bug localization examples, we found that the JIT compiler code was
included in the slice even with codegen dependencies turned off.
We are currently working on determining whether this is due to
legitimate dependencies in these computations or whether this 
is due to imprecision in our slicing algorithm.

\subsection{Evaluation Results: Summary}

The results from our experimental evaluation demonstrate the utility of
DCFGs and codegen dependencies for end-to-end analysis of dynamic code.
In particular, we show the following:
\begin{enumerate}
\item
  Backward analyses, augmented with
  the notion of codegen dependencies and extended to work over DCFGs, can
  effectively work back from specific behaviors in JIT-compiled code to the
  logic in the JIT compiler that resulted in the generation of that code.

\item
  Forward analyses, augmented with
  the notion of codegen dependencies and extended to work over DCFGs, can
  identify dynamic code that has been affected by environmental
  triggers in the course of dynamic code generation.

\item
  Our implementation of backward dynamic slicing is practical, in terms of
  analysis time; useful, in terms of the amount of code removed from the DCFG; and
  scalable, in terms of being able to handle analyses of significant size
  involving complex real-world software.
\end{enumerate}
Experiments with three widely-used dynamic analysis
toolkits that implement backward dynamic slicing.
show that these tools cannot
work back from observed behaviors in dynamically generated code to the relevant
code in the dynamic code generator.
Finally, for the identification trigger-based behaviors in
dynamic code, current tools based on stock symbolic execution
are not able to fully detect or explore all possible paths in the dynamic
code modifier, while current tools based on whole-system taint analysis are not able
to handle implicit information flows.

\section{Related Work}\label{sec:related-work}

There is a large body of literature on program slicing (e.g., see
\cite{weiser-slicing,ZGZ04,korel1997computation,SFB07,ranganath2007new}), 
but to the best of our knowledge all of this work focuses on static code.  
There is also a significant amount of research on various aspects of JIT
compilation, including
implementation \cite{Tracemonkey-JIT,Bolz09,Bebenita10,rompf2014surgical}
and taint analysis in the JITted code 
\cite{kerschbaumer2013information,rompf2014surgical,nair2008trishul,enck2014taintdroid}.
None of these works consider analyses that capture
dependencies between the JITted code and the code in the JIT compiler.

There is a lot of work on dependence and information flow analyses
(e.g., see \cite{NS-05,HTMF13,kang2011dta++}), 
but these typically do not consider end-to-end analysis of dynamic code.  
Several authors have discussed taint propagation in JIT-compiled code, but
focusing on taint propagation in just the application code rather
than on end-to-end analyses 
\cite{kerschbaumer2013information,rompf2014surgical,nair2008trishul,enck2014taintdroid}.
Whole-system analyses \cite{yin2007panorama,yin2010temu,henderson2014make,%
chow2004understanding,enck2014taintdroid}
focus on issues relating to dynamic taint propagation through the entire
computer system.  Such systems provide end-to-end analyses but
typically consider only explicit information flows ($\simeq$ data dependencies),
not implicit flows ($\simeq$ control dependencies); they are thus of limited use
for reasoning about behaviors, such as conditional dynamic code modification
(i.e., where the dynamic code generated may depend conditionally on input
and/or environmental values), which are common in applications such as JIT
compilers.

Korczynski and Yin discuss identifying code reuse/injections using whole-system
dynamic taint analysis \cite{korczynski2017capturing}.  While this work captures
codegen dependencies, it does not propose a program representation that can
capture the code structure for the different phases that arise during execution.
As a result, this approach is not suitable for analyses, such as program slicing,
that require information about the control flow structure of the code.

Dalla Preda {\em et al.} describe a notion of phases to characterize the semantics
of self-modifying code \cite{PGD15}.  The technical details of their work are very
different from ours.
Debray and Patel suggest using a notion similar to our codegen dependencies to identify
dynamic code unpackers \cite{DP10}, but focus exclusively on
code unpacking (e.g., do not consider applications such as bug localization
or environmental triggers), consider only relatively small malware samples,
and do not address the complexities of large systems such as V8.

There are a number of systems that reason about program behavior using dynamic 
analysis, and therefore are able to perform some kinds of analysis on dynamic
code; examples include PinPlay \cite{patil2010pinplay}, angr
\cite{shoshitaishvili2016state,stephens2016driller}, and 
 Triton \cite{SSTIC2015-Saudel-Salwan}.
 Our experiments indicate that these systems do not keep track of multiple versions 
 of code resulting from dynamic code modification, and as a result cannot fully
 capture all of the dependencies arising from runtime code changes.

Cai {\em et al.} \cite{cai2007certified} and Myreen \cite{myreen2010verified} 
discuss reasoning about dynamic code for the purposes of program verification 
using Hoare logic. We have not seen any implementations to apply their work 
towards modern software that utilizes dynamic code (i.e. a javascript engine). 
Furthermore, our work is more specific in that we seek to provide a program 
representation capable of representing dynamic code. 


\section{Conclusions}\label{sec:conclusions}

Dynamic code is ubiquitous in today's world
and raises its own set of security challenges.
Existing approaches to program analysis are not adequate for reasoning about
the behavior of dynamic code due to two main challenges: first, current program
representations do not adequately capture the
effects of changes due to dynamic code modification; second, they do not
account for dependencies that can arise due to the act of dynamic code
modification.
This paper discusses how these challenges can be addressed via
a program representation suitable for dynamic code as well as a new notion of
dependencies that can capture dependencies between the dynamic
code and the code that generated it.
Experiments with a prototype implementation of backwards dynamic slicing
based on these ideas show, on a number of real-world examples, that these ideas
make it possible to work back from the faulty code to the JIT compiler logic
that led to the generation of the faulty code.


%
\bibliography{ms}

%

\end{document}